\documentclass[twocolumn,showpacs,preprintnumbers,amsmath,amssymb,aps]{revtex4}

\usepackage{graphicx}
\usepackage{dcolumn}
\usepackage{bm}

\newcommand{\norm}[1]{\left\Vert#1\right\Vert}
\newcommand{\abs}[1]{\left\vert#1\right\vert}
\newcommand{\set}[1]{\left\{#1\right\}}
\newcommand{\Real}{\mathbb R}
\newcommand{\eps}{\varepsilon}
\newcommand{\To}{\longrightarrow}
\newcommand{\BX}{\mathbf{B}(X)}
\newcommand{\A}{\mathcal{A}}

\begin{document}


\title{Symmetry relating Gravity with Antigravity: A possible resolution of the
Cosmological Constant Problem?}

\author{Israel Quiros}

\email{israel@uclv.educ.u}

\affiliation{Universidad Central de Las Villas. Santa Clara. Cuba}

\date{\today}

\begin{abstract}
I discuss possible implications a symmetry relating gravity with
antigravity might have for smoothing out of the cosmological
constant puzzle. For this purpose, a very simple model with
spontaneous symmetry breaking is explored, that is based on
Einstein-Hilbert gravity with two self-interacting scalar fields.
The second (exotic) scalar particle with negative energy density,
could be interpreted, alternatively, as an antigravitating
particle with positive energy.
\end{abstract}



\maketitle

One of the most profound mysteries in fundamental physics is the
cosmological constant problem (see references \cite{lambdap} for
resent reviews on this subject). Recently there have been revealed
two faces of this problem: 1) Why is the cosmological constant so
small? and 2) Why is it comparable to the critical density of the
Universe precisely at present? In the present letter I will focus
on point 1) of the problem, leaving point 2) for further research.

The physical basis for the cosmological constant $\Lambda$ are the
zero-point vacuum fluctuations. The expectation value of the
energy-momentum tensor for vacuum can be written in the Lorentz
invariant form $<T_{ab}>_{vac}=(\Lambda/8\pi G)g_{ab}$, where $G$
is Newton's constant. It is divergent for both bosons and
fermions. Since bosons and fermions (of identical mass) contribute
equally but with opposite sign to the vacuum expectation of
physical quantities, supersymmetry was expected to account for the
(nearly) zero value of the cosmological constant, through an
accurate balance between bosons and fermions in Nature. However,
among other objections, the resulting scenario is not the one it
is expected to occur if a Universe with an early period of
inflation (large $\Lambda$) and a very small current value of
$\Lambda$ is to be described\cite{sahni}. Although other
mechanisms and principles, among them a running $\Lambda$ and the
anthropic principle\cite{lambdap,sahni}, have been invoked to
solve the cosmological constant (vacuum energy density) puzzle,
none of them have been able to give a definitive answer to this
question and the problem still remains a mystery.

At present, there is no known fundamental symmetry in Nature which
will set to zero the value of $\Lambda$ \cite{lambdap,sahni}.
Hence, the search for such a symmetry remains a challenge. In the
present letter I want to put forward the possibility that such a
fundamental symmetry could be the one under the interchange of
gravity and antigravity. The idea behind this possibility is that,
once this symmetry is assumed, to each gravitating standard model
particle (antiparticle), it corresponds an antigravitating
partner, whose contribution to the vacuum energy exactly cancels
gravitating particle's
contribution.\footnotemark\footnotetext{Gravitating and
antigravitating partners differ only in the sign of the mass
parameter, otherwise, in the character of their gravitational
interactions. All the other quantum numbers are identical.} This
can be visualized as if there were two distinct vacua: one
gravitating and other one that antigravitates so that, the
resulting "total" (averaged) vacuum does not gravitates at all.
The kind of symmetry I am proposing to account for the (nearly)
zero value of the cosmological constant, opens up the possibility
that such exotic entities like antigravitating objects, might
exist in Nature.\footnotemark\footnotetext{This would double the
number of existing standard model particles (antiparticles).}
Hence, why this kind of objects are not being observed in our
Universe? A similar question, this time for antimatter, has been
raised before. In this last case, a possible mechanism for
generating the desired amount of baryon asymmetry\cite{m-antim},
relies on three necessary (Sakharov's) conditions: i) Baryon
number non-conservation, ii) C and CP violation and iii)
Deviations from thermal equilibrium. In the same fashion, an
answer to the problem of gravitating-antigravitating matter
asymmetry could be approached. In this sense, one should expect
non-conservation of the charge associated with gravity-antigravity
symmetry.

In this letter I call gravity-antigravity transformations (G-aG
transformations for short) the following set of simultaneous
transformations: $G\rightarrow -G$ and $g_{ab}\rightarrow
-g_{ab}$. This means that, simultaneous with the interchange of
gravity and antigravity, interchange of time-like and space-like
domains is also required.\footnotemark\footnotetext{The
interchange of time-like and space-like domains, simultaneous with
the transformation $G\rightarrow -G$, is justified since, the last
transformation can be viewed as the interchange of Planck mass
squared and negative (tachyonic) Planck mass squared:
$M_{Pl}^2\rightarrow -M_{Pl}^2$.} It is straightforward noting
that, the purely gravity part of the Einstein-Hilbert action
$S=\int d^4x\sqrt{|g|} R/(16\pi G)$, is G-aG symmetric. Actually,
under $g_{ab}\rightarrow -g_{ab}$, the Ricci tensor is unchanged
$R_{ab}\rightarrow R_{ab}$, while the curvature scalar
$R\rightarrow -R$. The introduction of a $\Lambda$ term in the
above Eisntein-Hilbert action breaks this symmetry. This fact
hints at the possibility that, precisely, this kind of symmetry
could account for a zero value of the cosmological constant
$\Lambda$.

Since models with spontaneous symmetry breaking are relevant to
the cosmological constant problem \cite{sahni}, I will explore a
very simple model, that is based on general relativity plus
self-interacting scalar fields with symmetry breaking potentials,
as sources of gravity. The starting point will be the following
action which includes a single scalar field
$\phi$:\footnotemark\footnotetext{I chose the following signature
for the metric: ($-+++$).}

\begin{equation}
S=\int\frac{d^4x\sqrt{|g|}}{16\pi G}\{R-(\nabla\phi)^2-2V(\phi)\},
\label{1}
\end{equation}
where $V(\phi)$ is the self-interaction (symmetry breaking)
potential. This action respects reflection symmetry
$\phi\rightarrow -\phi$ if $V$ is an even function. However, it is
invariant under G-aG transformations, with the inclusion of
reflection, only if, simultaneously, $V(\phi)\rightarrow
-V(\phi)$. The last transformation, under reflection, restricts
the self-interaction potential to be an odd function of $\phi$.
Therefore, a very wide class of symmetry breaking potentials
(including the typical "Mexican hat" potential) is ruled out by
G-aG symmetry. In order to extend this symmetry to any kind of
potential, one can introduce a second self-interacting scalar
field $\bar\phi$, with the wrong sign of both kinetic and
potential energy terms, in the action (\ref{1}) and, at the same
time, to introduce the innocuous factor $\epsilon\equiv G/|G|$
($\epsilon=+1$ for gravity and $\epsilon=-1$ for antigravity) in
both kinetic terms for $\phi$ and $\bar\phi$. The improved action
is:

\begin{eqnarray}
S=\int\frac{d^4x\sqrt{|g|}}{16\pi\epsilon
|G|}\{R-\epsilon(\nabla\phi)^2-2V(\phi)\nonumber\\
+\epsilon(\nabla\bar\phi)^2+2V(\bar\phi)\}. \label{2}
\end{eqnarray}

Notice I kept the same symbol $V$ for the self-interacting
potential, meaning that the functional form of both $V(\phi)$ and
$V(\bar\phi)$ is the same. The fact that both kinetic and
potential energies of $\bar\phi$ enter with the wrong sign, means
that the energy density of the second scalar field
$\rho_{\bar\phi}=\epsilon(\nabla\bar\phi)^2/2-V(\bar\phi)$ is
negative if the potential $V$ is a positive definite function. A
second interpretation could be that, $\bar\phi$ has positive
energy
($\rho_{\bar\phi}\rightarrow\rho_{\bar\phi}^+=-\rho_{\bar\phi}>0$),
but it antigravitates ($\epsilon\rightarrow -\epsilon$). This
second interpretation is apparent if one realizes that, in the
right-hand side (RHS) of Einstein's equations, that are derivable
from (\ref{2}), one has the combination: $8\pi\epsilon |G|
(T_{ab}^\phi-T_{ab}^{\bar\phi})$, where the stress-energy tensor
for scalar field degrees of freedom is defined in the usual way
(except for the innocuous factor $\epsilon$):
$T_{ab}^\chi=\epsilon(\nabla_a\chi\nabla_b\chi-\frac{1}{2}g_{ab}(\nabla\chi)^2)
-g_{ab}V(\chi)$ ($\chi$ is the collective name for $\phi$ and
$\bar\phi$). Hence, one could hold the view that, the minus sign
in the second term of the RHS of Einstein's field equations, could
be absorbed into the innocuous factor $\epsilon$: $8\pi |G|
(\epsilon T_{ab}^\phi+(-\epsilon) T_{ab}^{\bar\phi})$. Any way,
$\bar\phi$ represents a kind of exotic particle, whose existence
could be justified only in quantum systems like the quantum
vacuum.\footnotemark\footnotetext{To learn about the recurrence of
negative energy densities in physics, see reference
\cite{negenerg}.}

The action (\ref{2}) is explicitly invariant under the (enhanced)
set of G-aG transformations:\footnotemark\footnotetext{If the
potential $V$ is an even function, then the reflection
$\phi\rightarrow -\phi$, $\bar\phi\rightarrow -\bar\phi$, is also
a symmetry of (\ref{2}).}

\begin{equation}
\epsilon\rightarrow -\epsilon,\;\; g_{ab}\rightarrow
-g_{ab},\;\;\phi\leftrightarrow\bar\phi. \label{3}
\end{equation}

A remarkable property of this model is that the Klein-Gordon
equations for both $\phi$ and $\bar\phi$

\begin{equation}
\Box\phi=\epsilon\frac{dV(\phi)}{d\phi},
\;\;\Box\bar\phi=\epsilon\frac{dV(\bar\phi)}{d\bar\phi}, \label{4}
\end{equation}
coincide. The consequence is that both fields will tend to run
down the potentials towards smaller energies. Therefore, if $V$
has global minima, both $\phi$ and $\bar\phi$ will tend to
approach one of these minima. This is, precisely, the key
ingredient in the present model to explain the small value of the
vacuum energy density, through weak violation of G-aG symmetry. To
illustrate this point, let us consider the "Mexican hat"
potential:

\begin{equation}
V(\chi)=V_0-\frac{\mu_\chi^2}{2}\chi^2+\frac{\lambda_\chi}{4}\chi^4,\;\;\lambda_\chi>0.
\label{5}
\end{equation}
The symmetric state $(\phi,\bar\phi)=(0,0)$ is unstable and the
system settles in one of the following ground states
$(\phi,\bar\phi)$:
$(\sqrt{\mu^2/\lambda},\sqrt{\bar\mu^2/\bar\lambda})$,
$(\sqrt{\mu^2/\lambda},-\sqrt{\bar\mu^2/\bar\lambda})$,
$(-\sqrt{\mu^2/\lambda},-\sqrt{\bar\mu^2/\bar\lambda})$,
$(-\sqrt{\mu^2/\lambda},\sqrt{\bar\mu^2/\bar\lambda})$. This means
that the reflection symmetry $\phi\rightarrow -\phi$,
$\bar\phi\rightarrow -\bar\phi$, inherent in theory (\ref{2}) with
potential (\ref{5}), is spontaneously broken. The immediate
consequence is that the stress-energy tensor of the vacuum of the
theory takes the Lorentz invariant form

\begin{equation}
T_{ab}=\frac{\Lambda}{8\pi |G|} g_{ab}, \label{6}
\end{equation}
where the cosmological constant
$\Lambda=((\mu^4/\lambda)-(\bar\mu^4/\bar\lambda))/4\epsilon$
($\mu\equiv\mu_\phi$, $\bar\mu\equiv\mu_{\bar\phi}$, etc.). Now it
is apparent that the resulting theory (with broken reflection
symmetry), that is given by the action $S=\int
d^4x\sqrt{|g|}\{R-2\Lambda\}/16\pi\epsilon |G|$, is invariant
under G-aG transformations (\ref{3}) only if
$(\mu,\lambda)=(\bar\mu,\bar\lambda)\Rightarrow\Lambda=0$. In
consequence, the small observed value $\Lambda\sim 10^{-47}
GeV^4$, denotes a weak violation of G-aG symmetry, that is due to
a degeneracy of reflection symmetry energy scales
$\alpha\equiv\mu^4/8\pi |G|\lambda$ and
$\bar\alpha\equiv\bar\mu^4/8\pi |G|\bar\lambda$.

Which physical mechanism is responsible for a small violation of
G-aG symmetry, is a question that could be clarified only once
exotic (in principle, antigravitating) fields like $\bar\phi$, are
built into a fundamental theory of the physical interactions,
including gravity.\footnotemark\footnotetext{The existing
(standard) model of the fundamental interactions, without the
inclusion of gravity, is unable to differentiate gravitating and
antigravitating particles.} In the absence of such fundamental
theory of the unified interactions, one might only conjecture on
the physical origin of the degeneracy in reflection symmetry
energy scales. In this regard, a possible origin of the
aforementioned degeneracy could be associated with the following
reasoning. It is an observational fact that, almost all real
(observable) standard model particles in Nature gravitate.
Therefore, their interactions with gravitating and antigravitating
vacuum particles are of different nature. The difference is
accentuated in the early stages of cosmic evolution, due to
strongest character of gravitational interactions for high-energy
scales, while, at present, the difference is very tiny, due to the
very weak intensity of gravitating effects at the low-energies
prevailing in the Universe. This will fit well a scenario, in
which a large value of $\Lambda$ is required to produce the due
amount of inflation in the early Universe, while a small current
value of $\Lambda$ will reproduce the present stage in the cosmic
evolution\cite{sahni}. A less physically inspired possibility is
that, during the course of the cosmic evolution, the relative
difference between reflection symmetry breaking scales is nearly
constant: $(\alpha-\bar\alpha)/\alpha\sim const.$ Therefore, for
larger (reflection) symmetry breaking scales, the cosmological
constant $\Lambda=2\pi |G|(\alpha-\bar\alpha)/\epsilon$ is larger.

Although the model explored in this letter is far from giving a
final answer to the cosmological constant puzzle and, besides, it
is incomplete in that it gives no explanation about a realistic
mechanism for present small violation of G-aG symmetry,
nevertheless, it hints at a possible connection between this
(would be) fundamental symmetry and the vacuum energy density. The
study of this symmetry could be relevant, also, to the
understanding of the role orbifold symmetry plays in
Randall-Sundrum (RS) brane models\cite{rs}. Actually, in RS
scenario $M_{Pl}^2\propto\int dy M_{Pl,5}^3$ ($y$ accounts for the
extra coordinate) so, symmetry under $y\rightarrow
-y\Leftrightarrow M_{Pl}^2\rightarrow -M_{Pl}^2$ ($G\rightarrow
-G$). Treatment of the second face of the cosmological constant
puzzle (see the introductory part of this letter), within the
present approach, requires of further research

I am grateful to the MES of Cuba for financial support of this
research.




\end{document}